\documentclass[pra,twocolumn,showpacs,amsmath,amssymb]{revtex4}

\usepackage{graphicx,color}

\begin{document}

\title{
	Many-body effects in a Bose-Fermi mixture	
}

\author{Kazuto Noda}
\email{noda@scphys.kyoto-u.ac.jp}
\affiliation{Department of Physics, Kyoto University, Kyoto 606-8502, Japan}
\author{Robert Peters}
\affiliation{Department of Physics, Kyoto University, Kyoto 606-8502, Japan}
\author{Norio Kawakami}
\affiliation{Department of Physics, Kyoto University, Kyoto 606-8502, Japan}
\author{Thomas Pruschke}
\affiliation{Institut f\"ur Theoretische Physik, Universit\"at G\"ottingen, G\"ottingen D-37077, Germany}

\begin{abstract}
We investigate many-body effects on a mixture of interacting bosons and fermions loaded in an optical lattice using a generalized dynamical mean field theory combined with the numerical renormalization group. We show that strong correlation effects emerge in the presence of bosonic superfluidity, leading to a renormalized peak structure near the Fermi level in the density of states for fermions. Remarkably, this kind of strong renormalization appears not only in the metallic phase but also in the insulating phases of fermions such as in the empty/filled band limit. A systematic analysis of the relation between the quasiparticle weight and the strength of superfluidity reveals that the renormalization effect is indeed caused by the boson degrees of freedom. It is found that such renormalization is also relevant to a supersolid phase consisting of a density wave ordering of fermions accompanied by bosonic superfluidity. This sheds light on the origin of the peak structure in the supersolid phase.

\end{abstract}

\maketitle

\section{Introduction}

Strongly interacting atoms in optical lattices have attracted much interest \cite{Bloch2008}. In these systems, one can tune the interaction strength and the lattice structure by controlling a magnetic field and the intensity of lasers. Due to such high controllability, the cold atom systems can be regarded as efficient simulators of quantum many-body physics. Several intriguing phenomena related to strong correlations were already observed experimentally, for example, a superfluid-Mott transition in bosonic systems \cite{Greiner2002}, a metal-Mott transition in fermionic systems \cite{Jordens2008,Schneider2008}, etc. 

Experimental research has already been extended to the topics which are not easy to investigate in conventional condensed matter physics. A typical example is a Bose-Fermi (BF) mixture realized in a harmonic trap \cite{Truscott2001,Stan2004,Inouye2004,Ospelkaus2006a,Zaccanti2006,Shin2008}, and also in optical lattices \cite{Gunter2006,Ospelkaus2006,Best2009a,Fukuhara2009,Wu2011,Sugawa2011,Park2011}. Rapid progress in these experiments has stimulated theoretical research on related topics \cite{Buhler2003,Mathey2004,Pollet2006,Pollet2008,Hebert2008,null2008,Titvinidze2008,Titvinidze2009,Orth2009,Orignac2010}; for example, the possibility of fascinating ground states such as a supersolid state (coexistence of bosonic superfluidity and density-wave ordering of fermions) has been proposed \cite{Buhler2003,Hebert2008,Titvinidze2008,Titvinidze2009,Orth2009}. In this context, Titvinidze et al. in Refs. \cite{Titvinidze2008,Titvinidze2009} pointed out that the density of states for fermions has an anomalous peak near the Fermi surface in the supersolid phase. It should be important and interesting to further clarify the many-body aspects of this structure, which naturally motivates us to provide a detailed analysis of it.

On the other hand, dynamical properties of the BF mixture systems have not been studied systematically. Recent rapid advances in probing dynamical properties of fermions via the rf spectroscopy make it possible to observe the single-particle excitation spectrum in a BCS-BEC crossover region \cite{Stewart2008,Gaebler2010,Perali2011}, the quasiparticle weight in a polaronic system (an extremely imbalance system) \cite{Schirotzek2009,Nascimb`ene2009} etc. These experimental developments would be also applied for the BF mixture systems in the near future, encouraging theoretical studies on dynamical properties of the BF mixture systems.

Motivated by these research activities, we investigate many-body effects on the BF Hubbard Hamiltonian with particular emphasis on its dynamical properties. We reveal unique features inherent in the BF mixture systems; the renormalization effect caused by the boson degrees of freedom gives rise to a peak structure near the Fermi level for the density of states both for metallic and insulating phases. A similar peak structure due to many-body effects appears even in the supersolid phase, which we will discuss in the following sections.
Our systematic study in this paper clearly explains that the origin of the peak structure is many-body effects induced by the interplay of the boson and fermion degrees of freedom.


This paper is organized as follows. In Sec. \ref{sec_hami}, we introduce the BF Hubbard model, and briefly explain the framework of a generalized version of dynamical mean field theory (DMFT), which extends the original fermionic DMFT to treat additional boson degrees of freedom. We make use of the numerical renormalization group (NRG) as an impurity solver of DMFT. In Sec. \ref{sec_withoutlong}, we reveal the renormalization effects in the presence of bosonic superfluidity. In the last part of this section, we shortly discuss how the many-body effects occur in the impurity model corresponding to the effective impurity model in the DMFT calculation. 
In Sec.\ref{sec_long} we discuss how the renormalization effects appear in a supersolid state. In Sec. \ref{sec_conclusion}, we summarize our results. 

\section{MODEL AND METHOD}
\label{sec_hami}

We consider a BF mixed system, which can be described by the following BF Hubbard Hamiltonian,
\begin{eqnarray}
H
&=&
H^{b}+H^{f}+H^{bf},
\\
H^{b}
&=&
-t^{b}\sum_{\langle i,j \rangle}b_{i}^{\dagger} b_{j}
-
\sum_{i}\mu^{b}n^{b}_{i}
+
\sum_{i}\frac{U^{b}}{2}n^{b}_{i}(n^{b}_{i}-1),
\nonumber
\\
\\
H^{f}
&=&
-t^{f}\sum_{\langle i,j \rangle}f_{i}^{\dagger} f_{j}
-
\sum_{i}\mu^{f}n^{f}_{i},
\\
H^{bf}
&=&
\sum_{i}U^{bf} n^{b}_{i} n^{f}_{i},
\end{eqnarray}
with $n^{b}_{i}=b^{\dagger}_{i}b_{i} \, (n^{f}_{i}=f^{\dagger}_{i}f_{i})$, where $b_{i} (f_{i})$ annihilates a boson (fermion) at site $i$. Here,  $t^{b}(t^{f})$ is the boson (fermion) transfer integral, $\mu^{b}(\mu^{f})$ the chemical potential for bosons (fermions) and $U^{b} (U^{bf})$ the on-site boson-boson (boson-fermion) interaction. Note that  $\langle i,j\rangle$ denotes the summation over the neighboring lattice sites. Note that we treat spinless (one-species) fermions in this article. The corresponding circumstance, where one species of fermions are mixed with bosons, has already been realized experimentally by properly selecting one of the hyperfine states in fermionic atoms (see \cite{Stan2004}).

To investigate the ground-state as well as dynamical properties of the system, we employ the DMFT. In order to treat the boson degrees of freedom, we use a generalized version of DMFT which is introduced in Ref. \cite{Titvinidze2009,Byczuk2009b}. In the generalized DMFT, the lattice model is mapped onto an effective impurity model embedded in an effective medium, as usually done in DMFT \cite{Metzner1989,Georges1996}. The Green's function is obtained via the self-consistent solution of this impurity model. This is why the DMFT exactly includes local quantum fluctuations, which cannot be taken into account by conventional mean-field approaches. The different point from the fermionic DMFT is that the bosonic superfluid order parameter, $\varphi = \langle b \rangle$, should be obtained in self-consistency steps. We perform the calculation using a semielliptic local density of states with a bandwidth $W=4t^{f}$ for the noninteracting system. In the following,  the half bandwidth $D=2t^{f}$ is used as a unit of energy.

In order to discuss many-body effects in the phases with and without a density-wave order, we introduce the corresponding effective impurity models. When we analyze the phases without a density wave order, we use the following generalized single impurity Anderson model as an effective impurity model,
\begin{eqnarray}
\label{eq_gsiam1}
H_{\rm GSIAM}
&=&
-zt^{b}(\varphi b^{\dagger} + \varphi^{\ast}b) + \frac{U^{b}}{2}n^{b}(n^{b} -1) - \mu^{b} n^{b} 
\nonumber \\
&&
-\mu^{f} n^{f} + \sum_k \{ \epsilon_k a_{k}^{\dagger}a_{k} + V_{k }(f^{\dagger} a_{k} + \mbox{h.c.})\}
\nonumber \\
&&
+ U^{bf}n^{b} n^{f},
\end{eqnarray}
where $z$ is the coordination number, $\varphi =\langle b \rangle$ the superfluid order parameter and $V_{k}$ the hybridization for fermions. 

For the phases with a density-wave order, we divide the bipartite lattice system into two sublattices \cite{Georges1996}. The corresponding Hamiltonian is
\begin{eqnarray}
\label{eq_gsiam2}
&&
H_{\rm GSIAM}
=
\nonumber
\\
&&
\sum_{\alpha=\pm 1}
\left[
-zt^{b}(\varphi_{\overline{\alpha}}b^{\dagger}_{\alpha}
+ 
\varphi_{\overline{\alpha}}^{\ast}b_{\alpha})
+
\frac{U^{b}}{2}n^{b}_{\alpha}(n^{b}_{\alpha}-1)
-
\mu^{b}_{\alpha}n^{b}_{\alpha}
\right.
\nonumber
\\
&&
-\mu^{f} n^{f}_{\alpha}
+
\sum_k
\{
\epsilon_k a_{k\alpha}^{\dagger}a_{k{\alpha}}
+
V_{k{\alpha}}(f^{\dagger}_{\alpha}a_{k{\alpha}}
    +     \mbox{h.c.})
\}
\nonumber
\\
&&
+
\left.
U^{bf}n^{b}_{\alpha}n^{f}_{\alpha}
\right],
\end{eqnarray}
where $\alpha=A,B$ represents the sublattice index ($\overline{\alpha}=A,B$ with $\overline{\alpha} \neq \alpha$). For this Hamiltonian, we perform single-site DMFT calculations for each sublattice structure.
In the following, we fix the parameters, $U^{b}=1.0,zt^{b}=0.05$.

We calculate the superfluid order parameter $\varphi$ and the self-energy $\Sigma^{bf}(\omega)$ self-consistently by employing the numerical renormalization group (NRG) \cite{Wilson1975,Bulla2008} as an impurity solver. NRG has the advantage in performing the high-accuracy calculation in the low energy region thanks to the logarithmic discretization of the conduction band. This method has been already extended to include boson degrees of freedom \cite{Titvinidze2008}, which are incorporated in the impurity Hamiltonian. This allows us to apply NRG with the same accuracy in the low energy region as the ordinary fermion case. 

We compute several thermodynamic quantities and the quasiparticle weight defined by
\begin{equation}
Z=\frac{1}{1-\left.d\Re\Sigma(\omega)/d\omega\right|_{\omega=0}},
\end{equation}
which is inversely proportional to the effective mass of fermions. This quantity represents how strong the correlation effect is.  We also calculate the density of states (DOS), $\rho(\omega)$, i.e. the single particle excitation spectra derived from the imaginary part of the Green's function.


\section{Many-body effects in normal phases of fermions}
\label{sec_withoutlong}

We first discuss many-body effects in the BF mixture system without a density-wave long-range order of fermions. We employ the effective impurity model described by Eq. (\ref{eq_gsiam1}) for generalized-DMFT calculations.

\subsection{Correlation effects in a metallic phase}

In order to figure out possible phases in the BF mixture system, we calculate several quantities such as the fermion filling $\langle n^{f} \rangle$, the boson filling $\langle n^{b} \rangle$, the bosonic superfluid order parameter $\varphi$ and the quasiparticle weight $Z$ as a function of  the chemical potential for bosons. In particular, the quasiparticle weight $Z$ can be used as a measure of correlation effects: for a free particle system $Z=1$, while for an extremely correlated system $Z \rightarrow 0$. 

\begin{figure}[htb]
\begin{tabular}{c}
\begin{minipage}{\hsize}
\begin{center}
\includegraphics[clip,width=0.8\linewidth]{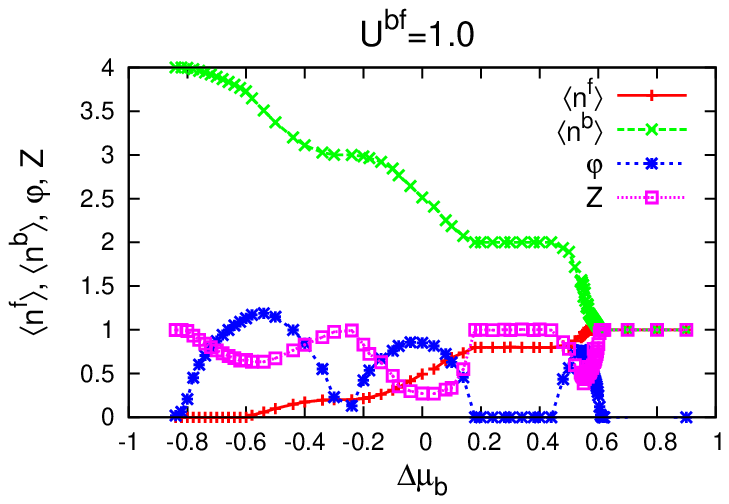}
\end{center}
\end{minipage}
\\
\begin{minipage}{\hsize}
\begin{center}
\includegraphics[clip,width=0.8\linewidth]{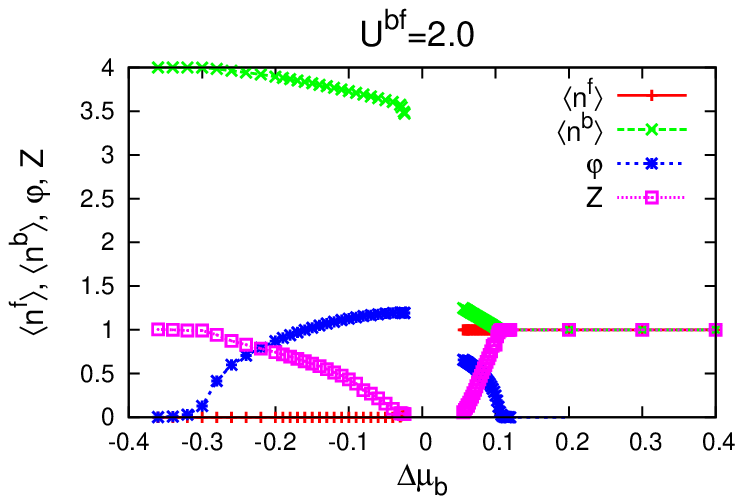}
\end{center}
\end{minipage}
\end{tabular}
\caption{(Color online) Fermion filling $\langle n^{f} \rangle$, boson filling $\langle n^{b} \rangle$, superfluid order parameter $\varphi$ and the quasiparticle weight $Z$ as a function of the chemical potential $\Delta \mu^{b}$ for bosons for fixed $U^{bf}=1.0$ (top) and $U^{bf}=2.0$ (bottom).}
\label{fig_ufb}
\end{figure}

The computed results are shown in Fig. \ref{fig_ufb}. Note that the origin of the chemical potential $\Delta \mu^b$ for bosons is defined so that it gives the fillings $\langle n^{f} \rangle=1/2,\langle n^{b} \rangle=5/2$  for $U^{bf}=1.0$ and $U^{bf}=2.0$. At non-integral fillings, the boson sector is always in a superfluid phase with finite  $\varphi$, while for $\langle n^{b}\rangle =1, 2, 3, 4$  (upper panel) and  $\langle n^{b} \rangle=1, 4$ (lower panel), it is in an insulating Mott phase with $\varphi=0$.  It is seen that there is no renormalization effect ($Z=1$) without superfluidity ($\varphi =0$).  On the other hand, in the presence of superfluidity, the quasiparticle weight $Z$ of fermions decreases from unity, implying that the renormalization of fermions occurs. Figure \ref{fig_ufb} also elucidates that the strength of the renormalization depends on the magnitude of superfluid order parameter $\varphi$; stronger renormalization (smaller $Z$) occurs for larger $\varphi$. These results certainly suggest that the many-body effects in the fermion sector are induced by the boson degrees of freedom via the boson-fermion interaction $U^{bf}$. 

We note here that the strong boson-fermion interaction may possibly induce orderings such as a density wave ordering both for the fermion and boson sectors. This kind of instability indeed appears around the region near $\Delta \mu_B \simeq 0$ in Fig. \ref{fig_ufb} ($U^{bf}=2.0$). The corresponding data for the physical quantities are lacking there because such density-wave ordering is not taken into account in the homogeneous DMFT calculations, so that there is no convergent solution. We will separately discuss the results for the density-wave state in Sec. \ref{sec_long}. 

\begin{figure}[htb]
\begin{center}
\includegraphics[clip,width=0.9\linewidth]{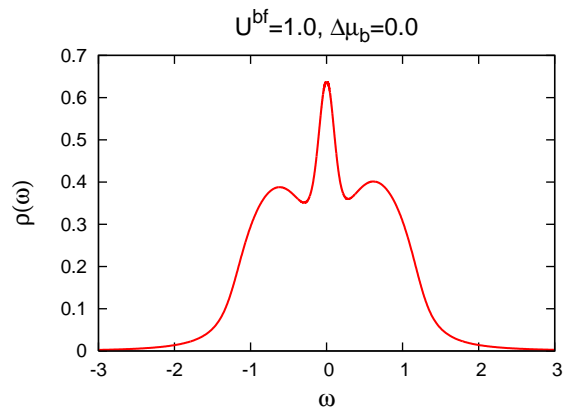}
\caption{(Color online) DOS for fermions with the boson-fermion interaction $U^{bf}=1.0$ and $\Delta\mu^{b}=0.0$, where the renormalization effects occur in the presence of bosonic superfluidity.}
\label{fig_ufb10dos}
\end{center}
\end{figure}

We now discuss how the renormalization effects appear in the DOS for fermions. In Fig.\ref{fig_ufb10dos}, we show a typical profile of the DOS for fermions in a metallic region with bosonic superfluidity at $U^{bf}=1.0$ and $\Delta \mu_{b}=0.0$ where fillings are $\langle n^{f} \rangle =1/2$ and $\langle n^{b} \rangle =5/2$. There is a sharp peak structure at the Fermi level due to the renormalization effects with bosonic superfluidity, which is one of the characteristic properties of the present BF mixture system. One may immediately notice that a similar peak structure is quite commonly observed in correlated electrons in condensed matter physics. Actually, there is a close relationship and a crucial difference between the present BF system and the electron systems; in both cases the strong renormalization of the DOS is caused by low-energy collective excitations, but in the former (latter) case the collective excitations come from electrons themselves (additional boson degrees of freedom). 

Therefore, we naturally expect that some intriguing aspects of correlation inherent in the BF systems should appear, which are not observed in correlated electron systems. Such examples can be indeed found in the empty/filled band limit of fermions in our case. We can see from Fig.\ref{fig_ufb} that the renormalization of fermions  occurs even in the extreme conditions for fermion filling: the empty band limit ($\langle n^{f} \rangle \sim 0$) or the filled-band limit ($\langle n^{f} \rangle \sim 1$). Since these cases provide unique aspects of the BF systems beyond ordinary electron systems, in the next subsection, we will give more detailed discussions on the many-body effects in the two limiting cases.

\subsection{Correlation effects in empty- and filled-band limit}

\subsubsection{empty-band limit}

\begin{figure}[t]
\begin{center}
\includegraphics[clip,width=0.7\linewidth]{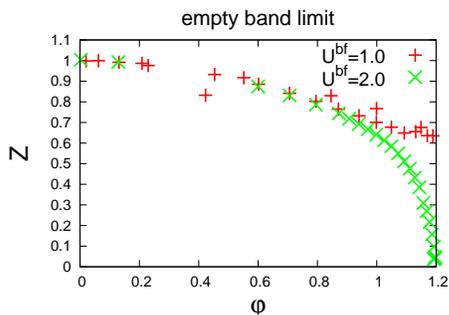}
\caption{(Color online) Quasiparticle weight $Z$ as a function of the superfluid order parameter $\varphi$ in the empty band limit. The boson-fermion interaction is fixed at $U^{bf}=1.0$ ($+$ points) and $U^{bf}=2.0$ ($\times$ points).}
\label{fig_zphiempty}
\end{center}
\end{figure}
\begin{figure}[t]
\begin{center}
\includegraphics[clip,width=1.0\linewidth]{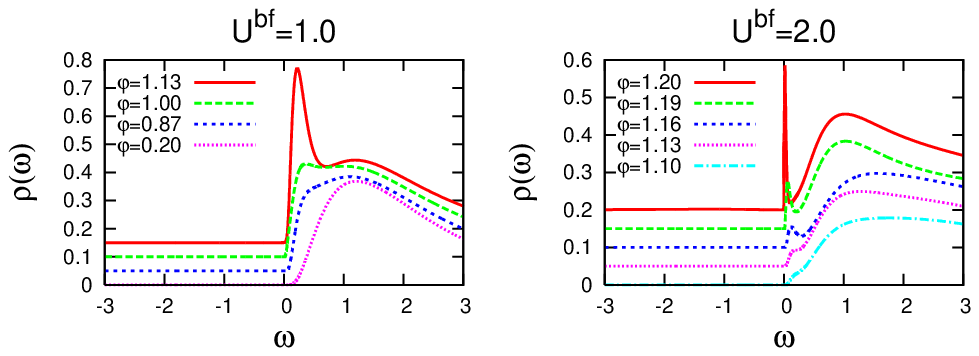}
\caption{(Color online) DOS as a function of the superfluid order parameter $\varphi$ at $U^{bf}=1.0$ (left panel) $2.0$ (right panel). The origin of the DOS is shifted for easy-to-see.}
\label{fig_dosempty}
\end{center}
\end{figure}

We first discuss the relation between the quasiparticle weight and the bosonic superfluid order parameter in the empty-band limit ($\langle n^{f} \rangle \sim 0$). The computed results are shown in Fig. \ref{fig_zphiempty}. It is seen that the evolution of the superfluid order parameter enhances the renormalization effects. This figure also indicates that the increase in the boson-fermion interaction $U^{bf}$ makes the renormalization effects stronger (smaller values of $Z$). For $U^{bf}=2.0$, the quasiparticle weight becomes almost zero around $\varphi \simeq 1.2$, which means that the quasiparticle mass of fermions becomes extremely heavy. This tendency may be related to instability toward the density-wave formation, as mentioned above.

In order to address how the renormalization affects dynamical properties, we calculate the DOS for fermions for several choices of $\varphi$. The results are plotted in Fig. \ref{fig_dosempty}.  Note that  only the particle-addition spectrum has finite values ($\omega>0$) because we cannot remove fermions from the system in this limiting case. In this figure the DOS has a peak near the Fermi level which becomes sharper as $\varphi$ becomes larger. This is consistent with the relation between the quasiparticle weight and the bosonic superfluidity mentioned above.

We note here that this kind of BF mixture in the empty-band limit has been already realized in recent experiments \cite{Wu2011,Park2011}, which is sometimes referred to as a "polaronic system" \cite{Schirotzek2009}. We hope that the above-mentioned renormalization effect could be observed experimentally in the near future.

\subsubsection{filled-band limit}

\begin{figure}[t]
\begin{center}
\includegraphics[clip,width=0.7\linewidth]{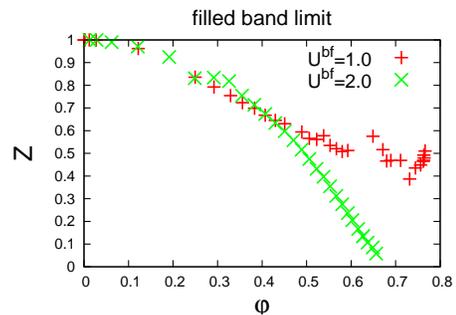}
\caption{(Color online) Quasiparticle weight $Z$ as a function of the superfluid order parameter $\varphi$ at empty band limit. The boson-fermion interaction is fixed at $U^{bf}=1.0$ ($+$ points) and $U^{bf}=2.0$ ($\times$ points).}
\label{fig_zphifull}
\end{center}
\end{figure}
\begin{figure}[t]
\begin{center}
\includegraphics[clip,width=1.0\linewidth]{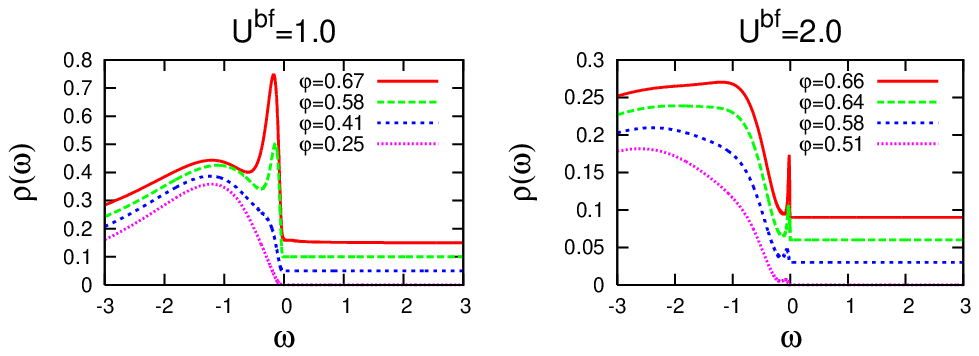}
\caption{(Color online) DOS as a function of the superfluid order parameter $\varphi$ at $U^{bf}=1.0$ (left panel) $2.0$ (right panel). The origin of the DOS is shifted for easy-to-see.}
\label{fig_dosfull}
\end{center}
\end{figure}

We next focus on the opposite extreme case, a filled-band limit, where the fermion filling is close to unity ($\langle n^{f} \rangle \sim 1$), while the boson sector is still at non-integer filling. In this limit, we can still tune the boson filling to control the amplitude of superfluidity. We plot the quasiparticle weight $Z$ as a function of the superfluid order parameter $\varphi$ in Fig. \ref{fig_zphifull}. Note here again that only the particle-removal spectrum has finite values ($\omega<0$) because we cannot add fermions into the system in this limiting case.

Figure \ref{fig_zphifull} suggests that the increase in the superfluid order parameter enhances the renormalization effects, resulting in smaller values of $Z$. Also, we can see that the boson-fermion interaction enhances the renormalization effects as for the empty band limit. We show typical examples of the DOS for several choices of $\varphi$  with fixed interactions in Fig. \ref{fig_dosfull}. With increasing $\varphi$, the peak becomes sharp in accordance with the corresponding behavior of the quasiparticle weight $Z$ as a function of $\varphi$. Therefore we conclude that the many-body effects induce the peak near the Fermi level in the DOS for fermion systems.

As mentioned above, in the filled-band limit, only the particle-removal spectrum can be observed. Experimentally, this limiting case may be more tractable than the empty-band limit if one could use the rf spectroscopy. We note that in this case, the renormalization effects should be discussed for a "hole" type quasiparticle.

\begin{figure}[t]
\begin{center}
\includegraphics[clip,width=0.9\linewidth]{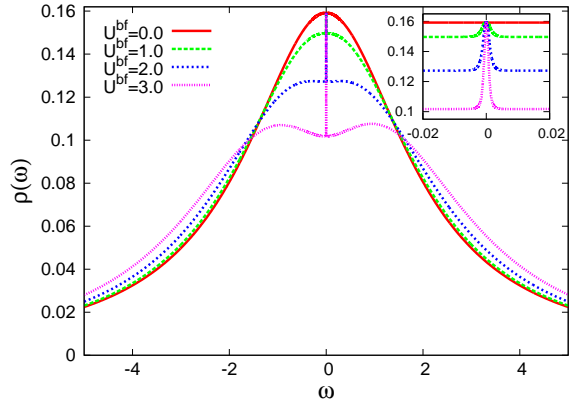}
\caption{(Color online) DOS for fermions in the impurity Hamiltonian with several boson-fermion interactions ($U^{bf}=0.0, 1.0, 2.0, 3.0$) in the presence of the bosonic superfluidity. A small peak structure on the Fermi surface disappears at $U=0.0$. Inset shows the enlarged view in the region around the Fermi surface.}
\label{fig_imppeak}
\end{center}
\end{figure}

Before concluding this section, we would like to briefly discuss the properties of the impurity Hamiltonian eq. (5) in order to exclude the possibility that the peak originates from the DMFT iteration process. We use NRG as an impurity solver with a constant density of states and a constant hybridization in order to extract general properties of the impurity Hamiltonian. Filling and interaction parameters are $\langle n^{f} \rangle=1/2,\langle n^{b} \rangle=5/2$, and $U^{b}=1.0$. Fig. \ref{fig_imppeak} shows the DOS for fermions with several boson-fermion interactions in the presence of the bosonic superfluidity. The DOS has two structures for finite boson-fermion interactions ($U^{bf}\neq0$). The main part of the excitation spectrum comes from the bare hybridization. On the other hand, the small peak on the Fermi surface only appears in the presence of finite boson-fermion interactions. We also confirm that the peak structure only appears in the presence of the bosonic superfluidity, which elucidates that the many-body effects occur in terms of boson degrees of freedom. These results are consistent with the DMFT calculations in the lattice system and support that the peak structure originates not from the DMFT iterations but from the interplay of the boson and fermion degrees of freedom.

\section{many-body effects in a supersolid phase}
\label{sec_long}

We now turn to the many-body effects in the phase with a density-wave order in the fermion sector. This case was already studied theoretically by Titvinidze et al. with DMFT \cite{Titvinidze2008,Titvinidze2009}, so we will perform complementary calculations to highlight the importance of many-body effects which was not addressed in the previous work. 

\begin{figure}[t]
\begin{center}
\includegraphics[clip,height=0.9\linewidth,angle=-90]{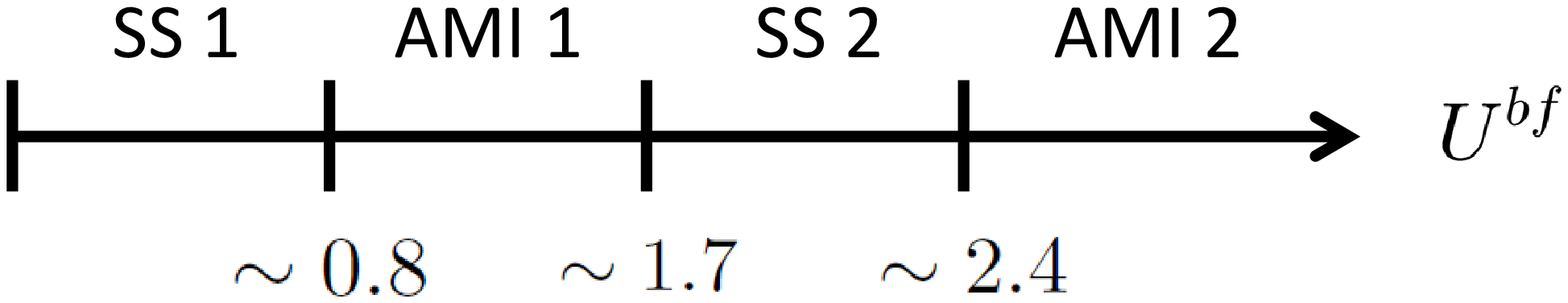}
\caption{Schematic phase diagram for the BF Hubbard model with fixed fillings $\langle n^{f} \rangle=1/2,\langle n^{b} \rangle=5/2$. SS means a supersolid phase and AMI an alternating Mott insulator.}
\label{fig_pd}
\end{center}
\end{figure}

To treat a density-wave order, we employ Eq.(\ref{eq_gsiam2}) as an effective impurity model, which allows us to treat the sublattice symmetry breaking. Here, we present the results for fixed fillings $\langle n^{f}\rangle=1/2$ and $ \langle n^{b}\rangle=5/2$, where we define $\langle n^{f(b)}\rangle =1/2\sum_{\alpha=A,B}n^{f(b)}_{\alpha}$. 
We briefly summarize the phase diagram with a density-wave order of fermions \cite{Titvinidze2008,Titvinidze2009}. As far as $U^{bf} \neq 0$ , there always exists the difference in the particle density between two sublattices, $\Delta N^{b(f)}=1/2|n^{b(f)}_{\alpha}-n^{b(f)}_{\overline{\alpha}}| \neq 0$, signaling the stability of a density wave state. The density wave state is referred to as an alternating Mott insulator (AMI) for bosons and as a "charge" density wave (CDW) for fermions, where "charge" is used following the tradition in the solid state physics. Furthermore, for $\Delta N^{b} \neq m+1/2$ ($m$ is integer), the boson sector favors a superfluid phase because the commensurability condition is not satisfied. Therefore, in the presence of bosonic superfluidity, the BF mixture system becomes a "supersolid" (SS) phase \cite{Titvinidze2009}, which is a main topic in this section.

\begin{figure}[t]
\begin{center}
\includegraphics[clip,width=0.8\linewidth]{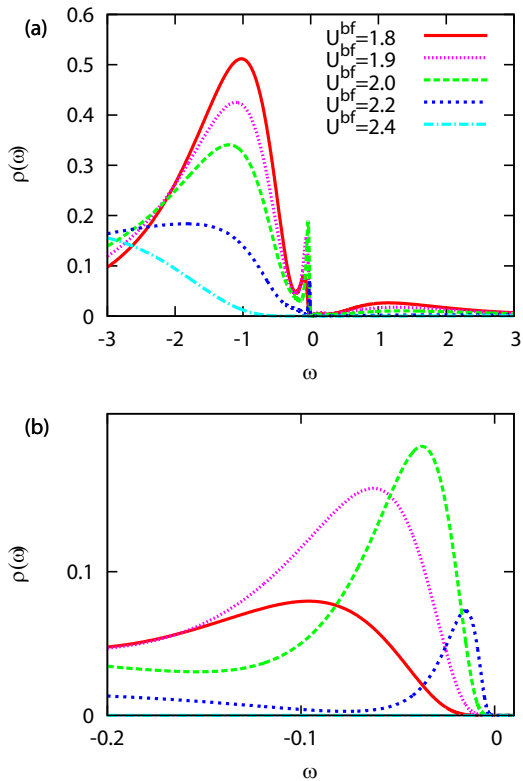}
\caption{(Color online) (a) DOS for $U^{bf}=1.8,1.9,2.0,2.2,2.4$. (b) DOS enlarged around the Fermi level.}
\label{fig_dos}
\end{center}
\end{figure}

In order to focus on the essential points, we here fix the boson-boson interaction $U^{b}=1.0$. In this condition, we end up with the phase diagram as a function of $U^{bf}$, as shown schematically in Fig.\ref{fig_pd}. The physical properties in this phase diagram are essentially the same as those obtained by Titvinidze et al. \cite{Titvinidze2009} for different fillings $\langle n^{f} \rangle =1/2$ and   $\langle n^{f} \rangle =3/2$. Interestingly, they pointed out that the DOS for fermions has a peak structure near the Fermi level in the SS phase. 
We demonstrate here  that this peak is caused by the many-body effects.

To address the above point, we focus on the supersolid SS2 phase where the correlation effects are observed more clearly than the SS1 phase. We show the DOS in the SS2 phase in Fig. \ref{fig_dos}. It is seen that the DOS has two characteristic structures, as expected. In the high energy region, there is a hump structure which comes from the mean-field type effect of $U^{bf}$. This exists both in the metallic and insulating phases. On the other hand, in the low-energy region, there is a sharp peak near the Fermi level, which only emerges in the supersolid phase, as already found  by Titvinidze et al. for different fillings \cite{Titvinidze2009}.

\begin{figure}[tb]
\begin{center}
\includegraphics[clip,width=0.85\linewidth]{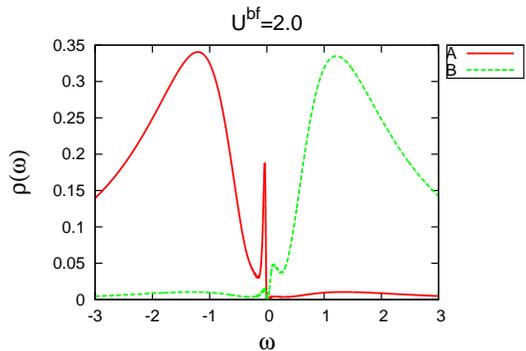}
\caption{(Color online) DOS for the A and B sublattices at $U^{bf}=2.0$.}
\label{fig_dos2}
\end{center}
\end{figure}

We now provide evidence that this peak indeed originates from the correlation effects due to boson degrees of freedom. We show the DOS for different choices of the interaction $U^{bf}$ in Fig.\ref{fig_dos}. With increasing $U^{bf}$, the weight of the peak initially increases and then decreases, as seen in the lower panel of Fig.\ref{fig_dos}. This behavior is similar to the one observed between the quasiparticle weight and the bosonic superfluidity, so that it is consistent with the results obtained in the previous section. In order to check the sublattice-dependence, we show the DOS for each sublattice in Fig. \ref{fig_dos2}. In both cases, the DOS has a peak near the Fermi level, suggesting that not only the A- but also B-sublattice has the anomalous peak, which was not obvious in the previous study \cite{Titvinidze2008,Titvinidze2009}. Note that the shape of DOS is quite different from each other. In the supersolid phase, the occupation number of fermions at each site is alternating between two sublattices: if $\Delta N^{f}$ is close to full ($ \langle n^{f}_{A} \rangle \sim 1$), the B-sublattice remains almost empty ($\langle n^{f}_{B}  \rangle \sim 0$), and vice versa. Note that the condition for each sublattice approximately corresponds to the empty- and filled band limit discussed in the previous section, and therefore the corresponding DOS indeed exhibits analogous properties discussed in Figs. \ref{fig_dosempty} and \ref{fig_dosfull}. Therefore, we can say that the emergent peak structure in the supersolid phase is a fingerprint of many-body effects inherent in the BF mixture system.

\begin{figure}[t]
\begin{center}
\includegraphics[clip,width=0.8\linewidth]{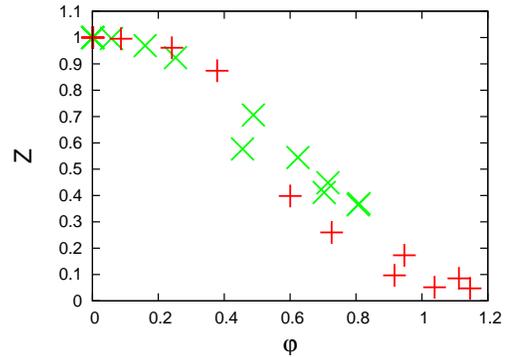}
\caption{(Color online) Quasiparticle weight $Z$ as a function of superfluid order parameter $\varphi$ for several values of interaction $1.6< U^{bf}< 2.5$. Fillings are fixed at $\langle n^{f} \rangle = 1/2, \langle n^{b} \rangle =5/2$. $+,\times$ correspond to the A, B sublattice in Fig.\ref{fig_dos2}.}
\label{fig_zphi}
\end{center}
\end{figure}

To further confirm our statement,  we show the relation between the quasiparticle weight $Z$ and the superfluid order parameter $\varphi$ in Fig. \ref{fig_zphi}.  The data include the calculations performed for several different interactions ($1.6 < U^{bf} < 2.5$) with fixed fillings $\langle n^{f} \rangle=1/2,\langle n^{b} \rangle=2/5$. It is seen that the increase of $\varphi$ enhances the renormalization effects (i.e. smaller $Z$), whereas there is no correlation effect (i.e. $Z=1$) in the absence of the bosonic superfluidity $\varphi=0$. This behavior is consistent with that for the metallic phase in the previous section. Therefore we confirm that the renormalization effects in the supersolid phase are induced by the boson degrees of freedom.

\section{CONCLUSION}
\label{sec_conclusion}

We have theoretically investigated a mixture of bosons and fermions loaded in an optical lattice using a generalized DMFT combined with NRG. We have revealed that strong correlation effects emerge in the fermion sector in the presence of bosonic superfluidity regardless of whether the system is metallic or insulating. This conclusion has been drawn via a systematic study of the close relationship between the renormalization factor and the magnitude of superfluidity. 

This kind of boson-driven renormalization effect gives rise to the characteristic peak structure in the low-energy region of the DOS. The formation of such a many-body peak is similar to that expected for the ordinary Fermi liquid, but there is a crucial difference between them. While in the ordinary Fermi liquid, the renormalization is caused by the low-energy excitations of fermions themselves, in the present mixture system, it is driven by boson degrees of freedom. Therefore, the latter brings about some unique correlation effects inherent in the BF mixture systems: for example, the strong renormalization appears even in the empty/filled band limit of fermions. Since the empty band limit was already realized in recent experiments, we hope that many body effects proposed here will be observed in the near future.



We have confirmed the appearance of the peak structure in the impurity model and concluded that the many-body effects occur in the presence of the bosonic superfluidity even in the impurity Hamiltonian. This implies that the peak structure originates from the boson degrees of freedom, not from DMFT iterations. In the preceding work [25] an instability towards the phase separation was pointed out as a possible mechanism for the peak-formation. Although in the present calculation, we have not encountered such a tendency, more detailed analyses should be necessary to figure out the relationship between our scenario and the previous one.

Although the calculation in this paper has been done at $T=0$, we expect that such an anomalous peak structure in the DOS can be observed  with the rf-spectroscopy experiments \cite{Stewart2008,Schirotzek2009,Gaebler2010,Perali2011} at sufficiently low temperatures where fermions and bosons are both in the quantum degenerate regime. The emergence of such a peak structure should be a fingerprint of the many-body effects inherent in the BF mixture.

\begin{acknowledgments}
We thank K. Inaba and A. Koga for stimulating discussions.  This work is supported by KAKENHI (Nos. 21540359, 20102008) and the Global COE Program "The Next Generation of Physics, Spun from Universality and Emergence" from MEXT of Japan. KN thanks Japan Society for the Promotion of Science (JSPS) for Research Fellowships for Young Scientists. RP thanks JSPS and the Alexander von Humboldt-Foundation. NK is supported by JSPS through its FIRST Program. TP also gratefully acknowledges support by JSPS through the Bridge program. Part of the computations was done at the Supercomputer Center at the Institute for Solid State Physics, University of Tokyo and Yukawa Institute Computer Facility.

\end{acknowledgments}


\begin{thebibliography}{37}
\expandafter\ifx\csname natexlab\endcsname\relax\def\natexlab#1{#1}\fi
\expandafter\ifx\csname bibnamefont\endcsname\relax
  \def\bibnamefont#1{#1}\fi
\expandafter\ifx\csname bibfnamefont\endcsname\relax
  \def\bibfnamefont#1{#1}\fi
\expandafter\ifx\csname citenamefont\endcsname\relax
  \def\citenamefont#1{#1}\fi
\expandafter\ifx\csname url\endcsname\relax
  \def\url#1{\texttt{#1}}\fi
\providecommand{\bibinfo}[2]{#2}
\providecommand{\eprint}[2][]{\url{#2}}

\bibitem[{\citenamefont{Bloch et~al.}(2008)\citenamefont{Bloch, Dalibard, and
  Zwerger}}]{Bloch2008}
\bibinfo{author}{\bibfnamefont{I.}~\bibnamefont{Bloch}},
  \bibinfo{author}{\bibfnamefont{J.}~\bibnamefont{Dalibard}}, \bibnamefont{and}
  \bibinfo{author}{\bibfnamefont{W.}~\bibnamefont{Zwerger}},
  \bibinfo{journal}{Rev. Mod. Phys.} \textbf{\bibinfo{volume}{80}},
  \bibinfo{pages}{885} (\bibinfo{year}{2008}).

\bibitem[{\citenamefont{Greiner et~al.}(2002)\citenamefont{Greiner, Mandel,
  Esslinger, Hansch, and Bloch}}]{Greiner2002}
\bibinfo{author}{\bibfnamefont{M.}~\bibnamefont{Greiner}},
  \bibinfo{author}{\bibfnamefont{O.}~\bibnamefont{Mandel}},
  \bibinfo{author}{\bibfnamefont{T.}~\bibnamefont{Esslinger}},
  \bibinfo{author}{\bibfnamefont{T.~W.} \bibnamefont{Hansch}},
  \bibnamefont{and} \bibinfo{author}{\bibfnamefont{I.}~\bibnamefont{Bloch}},
  \bibinfo{journal}{Nature} \textbf{\bibinfo{volume}{415}}, \bibinfo{pages}{39}
  (\bibinfo{year}{2002}).

\bibitem[{\citenamefont{Jordens et~al.}(2008)\citenamefont{Jordens, Strohmaier,
  Gunter, Moritz, and Esslinger}}]{Jordens2008}
\bibinfo{author}{\bibfnamefont{R.}~\bibnamefont{Jordens}},
  \bibinfo{author}{\bibfnamefont{N.}~\bibnamefont{Strohmaier}},
  \bibinfo{author}{\bibfnamefont{K.}~\bibnamefont{Gunter}},
  \bibinfo{author}{\bibfnamefont{H.}~\bibnamefont{Moritz}}, \bibnamefont{and}
  \bibinfo{author}{\bibfnamefont{T.}~\bibnamefont{Esslinger}},
  \bibinfo{journal}{Nature} \textbf{\bibinfo{volume}{455}},
  \bibinfo{pages}{204} (\bibinfo{year}{2008}).

\bibitem[{\citenamefont{Schneider et~al.}(2008)\citenamefont{Schneider,
  Hackerm\"uller, Will, Best, Bloch, Costi, Helmes, Rasch, and
  Rosch}}]{Schneider2008}
\bibinfo{author}{\bibfnamefont{U.}~\bibnamefont{Schneider}},
  \bibinfo{author}{\bibfnamefont{L.}~\bibnamefont{Hackerm\"uller}},
  \bibinfo{author}{\bibfnamefont{S.}~\bibnamefont{Will}},
  \bibinfo{author}{\bibfnamefont{T.}~\bibnamefont{Best}},
  \bibinfo{author}{\bibfnamefont{I.}~\bibnamefont{Bloch}},
  \bibinfo{author}{\bibfnamefont{T.~A.} \bibnamefont{Costi}},
  \bibinfo{author}{\bibfnamefont{R.~W.} \bibnamefont{Helmes}},
  \bibinfo{author}{\bibfnamefont{D.}~\bibnamefont{Rasch}}, \bibnamefont{and}
  \bibinfo{author}{\bibfnamefont{A.}~\bibnamefont{Rosch}},
  \bibinfo{journal}{Science} \textbf{\bibinfo{volume}{322}},
  \bibinfo{pages}{1520} (\bibinfo{year}{2008}).

\bibitem[{\citenamefont{Truscott et~al.}(2001)\citenamefont{Truscott, Strecker,
  McAlexander, Partridge, and Hulet}}]{Truscott2001}
\bibinfo{author}{\bibfnamefont{A.~G.} \bibnamefont{Truscott}},
  \bibinfo{author}{\bibfnamefont{K.~E.} \bibnamefont{Strecker}},
  \bibinfo{author}{\bibfnamefont{W.~I.} \bibnamefont{McAlexander}},
  \bibinfo{author}{\bibfnamefont{G.~B.} \bibnamefont{Partridge}},
  \bibnamefont{and} \bibinfo{author}{\bibfnamefont{R.~G.} \bibnamefont{Hulet}},
  \bibinfo{journal}{Science} \textbf{\bibinfo{volume}{291}},
  \bibinfo{pages}{2570} (\bibinfo{year}{2001}).

\bibitem[{\citenamefont{Stan et~al.}(2004)\citenamefont{Stan, Zwierlein,
  Schunck, Raupach, and Ketterle}}]{Stan2004}
\bibinfo{author}{\bibfnamefont{C.~A.} \bibnamefont{Stan}},
  \bibinfo{author}{\bibfnamefont{M.~W.} \bibnamefont{Zwierlein}},
  \bibinfo{author}{\bibfnamefont{C.~H.} \bibnamefont{Schunck}},
  \bibinfo{author}{\bibfnamefont{S.~M.~F.} \bibnamefont{Raupach}},
  \bibnamefont{and} \bibinfo{author}{\bibfnamefont{W.}~\bibnamefont{Ketterle}},
  \bibinfo{journal}{Phys. Rev. Lett.} \textbf{\bibinfo{volume}{93}},
  \bibinfo{pages}{143001} (\bibinfo{year}{2004}).

\bibitem[{\citenamefont{Inouye et~al.}(2004)\citenamefont{Inouye, Goldwin,
  Olsen, Ticknor, Bohn, and Jin}}]{Inouye2004}
\bibinfo{author}{\bibfnamefont{S.}~\bibnamefont{Inouye}},
  \bibinfo{author}{\bibfnamefont{J.}~\bibnamefont{Goldwin}},
  \bibinfo{author}{\bibfnamefont{M.~L.} \bibnamefont{Olsen}},
  \bibinfo{author}{\bibfnamefont{C.}~\bibnamefont{Ticknor}},
  \bibinfo{author}{\bibfnamefont{J.~L.} \bibnamefont{Bohn}}, \bibnamefont{and}
  \bibinfo{author}{\bibfnamefont{D.~S.} \bibnamefont{Jin}},
  \bibinfo{journal}{Phys. Rev. Lett.} \textbf{\bibinfo{volume}{93}},
  \bibinfo{pages}{183201} (\bibinfo{year}{2004}).

\bibitem[{\citenamefont{Ospelkaus
  et~al.}(2006{\natexlab{a}})\citenamefont{Ospelkaus, Ospelkaus, Humbert,
  Sengstock, and Bongs}}]{Ospelkaus2006a}
\bibinfo{author}{\bibfnamefont{S.}~\bibnamefont{Ospelkaus}},
  \bibinfo{author}{\bibfnamefont{C.}~\bibnamefont{Ospelkaus}},
  \bibinfo{author}{\bibfnamefont{L.}~\bibnamefont{Humbert}},
  \bibinfo{author}{\bibfnamefont{K.}~\bibnamefont{Sengstock}},
  \bibnamefont{and} \bibinfo{author}{\bibfnamefont{K.}~\bibnamefont{Bongs}},
  \bibinfo{journal}{Phys. Rev. Lett.} \textbf{\bibinfo{volume}{97}},
  \bibinfo{eid}{120403} (\bibinfo{year}{2006}{\natexlab{a}}).
  
\bibitem[{\citenamefont{Zaccanti et~al.}(2006)\citenamefont{Zaccanti, D'Errico,
  Ferlaino, Roati, Inguscio, and Modugno}}]{Zaccanti2006}
\bibinfo{author}{\bibfnamefont{M.}~\bibnamefont{Zaccanti}},
  \bibinfo{author}{\bibfnamefont{C.}~\bibnamefont{D'Errico}},
  \bibinfo{author}{\bibfnamefont{F.}~\bibnamefont{Ferlaino}},
  \bibinfo{author}{\bibfnamefont{G.}~\bibnamefont{Roati}},
  \bibinfo{author}{\bibfnamefont{M.}~\bibnamefont{Inguscio}}, \bibnamefont{and}
  \bibinfo{author}{\bibfnamefont{G.}~\bibnamefont{Modugno}},
  \bibinfo{journal}{Phys. Rev. A} \textbf{\bibinfo{volume}{74}},
  \bibinfo{eid}{041605} (\bibinfo{year}{2006}).

\bibitem[{\citenamefont{Shin et~al.}(2008)\citenamefont{Shin, Schirotzek,
  Schunck, and Ketterle}}]{Shin2008}
\bibinfo{author}{\bibfnamefont{Y.-i.} \bibnamefont{Shin}},
  \bibinfo{author}{\bibfnamefont{A.}~\bibnamefont{Schirotzek}},
  \bibinfo{author}{\bibfnamefont{C.~H.} \bibnamefont{Schunck}},
  \bibnamefont{and} \bibinfo{author}{\bibfnamefont{W.}~\bibnamefont{Ketterle}},
  \bibinfo{journal}{Phys. Rev. Lett.} \textbf{\bibinfo{volume}{101}},
  \bibinfo{pages}{070404} (\bibinfo{year}{2008}).

\bibitem[{\citenamefont{Gunter et~al.}(2006)\citenamefont{Gunter, Stoferle,
  Moritz, Kohl, and Esslinger}}]{Gunter2006}
\bibinfo{author}{\bibfnamefont{K.}~\bibnamefont{Gunter}},
  \bibinfo{author}{\bibfnamefont{T.}~\bibnamefont{Stoferle}},
  \bibinfo{author}{\bibfnamefont{H.}~\bibnamefont{Moritz}},
  \bibinfo{author}{\bibfnamefont{M.}~\bibnamefont{Kohl}}, \bibnamefont{and}
  \bibinfo{author}{\bibfnamefont{T.}~\bibnamefont{Esslinger}},
  \bibinfo{journal}{Phys. Rev. Lett.} \textbf{\bibinfo{volume}{96}},
  \bibinfo{eid}{180402} (\bibinfo{year}{2006}).

\bibitem[{\citenamefont{Ospelkaus
  et~al.}(2006{\natexlab{b}})\citenamefont{Ospelkaus, Ospelkaus, Wille, Succo,
  Ernst, Sengstock, and Bongs}}]{Ospelkaus2006}
\bibinfo{author}{\bibfnamefont{S.}~\bibnamefont{Ospelkaus}},
  \bibinfo{author}{\bibfnamefont{C.}~\bibnamefont{Ospelkaus}},
  \bibinfo{author}{\bibfnamefont{O.}~\bibnamefont{Wille}},
  \bibinfo{author}{\bibfnamefont{M.}~\bibnamefont{Succo}},
  \bibinfo{author}{\bibfnamefont{P.}~\bibnamefont{Ernst}},
  \bibinfo{author}{\bibfnamefont{K.}~\bibnamefont{Sengstock}},
  \bibnamefont{and} \bibinfo{author}{\bibfnamefont{K.}~\bibnamefont{Bongs}},
  \bibinfo{journal}{Phys. Rev. Lett.} \textbf{\bibinfo{volume}{96}},
  \bibinfo{eid}{180403} (\bibinfo{year}{2006}{\natexlab{b}}).

\bibitem[{\citenamefont{Best et~al.}(2009)\citenamefont{Best, Will, Schneider,
  Hackerm\"uller, van Oosten, Bloch, and L\"uhmann}}]{Best2009a}
\bibinfo{author}{\bibfnamefont{T.}~\bibnamefont{Best}},
  \bibinfo{author}{\bibfnamefont{S.}~\bibnamefont{Will}},
  \bibinfo{author}{\bibfnamefont{U.}~\bibnamefont{Schneider}},
  \bibinfo{author}{\bibfnamefont{L.}~\bibnamefont{Hackerm\"uller}},
  \bibinfo{author}{\bibfnamefont{D.}~\bibnamefont{van Oosten}},
  \bibinfo{author}{\bibfnamefont{I.}~\bibnamefont{Bloch}}, \bibnamefont{and}
  \bibinfo{author}{\bibfnamefont{D.-S.} \bibnamefont{L\"uhmann}},
  \bibinfo{journal}{Phys. Rev. Lett.} \textbf{\bibinfo{volume}{102}},
  \bibinfo{pages}{030408} (\bibinfo{year}{2009}).

\bibitem[{\citenamefont{Fukuhara et~al.}(2009)\citenamefont{Fukuhara, Sugawa,
  Takasu, and Takahashi}}]{Fukuhara2009}
\bibinfo{author}{\bibfnamefont{T.}~\bibnamefont{Fukuhara}},
  \bibinfo{author}{\bibfnamefont{S.}~\bibnamefont{Sugawa}},
  \bibinfo{author}{\bibfnamefont{Y.}~\bibnamefont{Takasu}}, \bibnamefont{and}
  \bibinfo{author}{\bibfnamefont{Y.}~\bibnamefont{Takahashi}},
  \bibinfo{journal}{Phys. Rev. A} \textbf{\bibinfo{volume}{79}},
  \bibinfo{pages}{021601} (\bibinfo{year}{2009}).

\bibitem[{\citenamefont{Sugawa et~al.}(2011)\citenamefont{Sugawa, Inaba, Taie,
  Yamazaki, Yamashita, and Takahashi}}]{Sugawa2011}
\bibinfo{author}{\bibfnamefont{S.}~\bibnamefont{Sugawa}},
  \bibinfo{author}{\bibfnamefont{K.}~\bibnamefont{Inaba}},
  \bibinfo{author}{\bibfnamefont{S.}~\bibnamefont{Taie}},
  \bibinfo{author}{\bibfnamefont{R.}~\bibnamefont{Yamazaki}},
  \bibinfo{author}{\bibfnamefont{M.}~\bibnamefont{Yamashita}},
  \bibnamefont{and}
  \bibinfo{author}{\bibfnamefont{Y.}~\bibnamefont{Takahashi}},
  \bibinfo{journal}{Nature Phys.} \textbf{\bibinfo{volume}{7}},
  \bibinfo{pages}{642} (\bibinfo{year}{2011}).

\bibitem[{\citenamefont{Wu et~al.}(2011)\citenamefont{Wu, Santiago, Park,
  Ahmadi, and Zwierlein}}]{Wu2011}
\bibinfo{author}{\bibfnamefont{C.-H.} \bibnamefont{Wu}},
  \bibinfo{author}{\bibfnamefont{I.}~\bibnamefont{Santiago}},
  \bibinfo{author}{\bibfnamefont{J.~W.} \bibnamefont{Park}},
  \bibinfo{author}{\bibfnamefont{P.}~\bibnamefont{Ahmadi}}, \bibnamefont{and}
  \bibinfo{author}{\bibfnamefont{M.~W.} \bibnamefont{Zwierlein}},
  \bibinfo{journal}{Phys. Rev. A} \textbf{\bibinfo{volume}{84}},
  \bibinfo{pages}{011601} (\bibinfo{year}{2011}).

\bibitem[{\citenamefont{{Park} et~al.}(2011)\citenamefont{{Park}, {Wu},
  {Santiago}, {Tiecke}, {Ahmadi}, and {Zwierlein}}}]{Park2011}
\bibinfo{author}{\bibfnamefont{J.~W.} \bibnamefont{{Park}}},
  \bibinfo{author}{\bibfnamefont{C.-H.} \bibnamefont{{Wu}}},
  \bibinfo{author}{\bibfnamefont{I.}~\bibnamefont{{Santiago}}},
  \bibinfo{author}{\bibfnamefont{T.~G.} \bibnamefont{{Tiecke}}},
  \bibinfo{author}{\bibfnamefont{P.}~\bibnamefont{{Ahmadi}}}, \bibnamefont{and}
  \bibinfo{author}{\bibfnamefont{M.~W.} \bibnamefont{{Zwierlein}}},
  \bibinfo{journal}{ArXiv e-prints}  (\bibinfo{year}{2011}),
  \eprint{1110.4552}.

\bibitem[{\citenamefont{Buchler and Blatter}(2003)}]{Buhler2003}
\bibinfo{author}{\bibfnamefont{H.~P.} \bibnamefont{Buchler}}
  \bibnamefont{and} \bibinfo{author}{\bibfnamefont{G.}~\bibnamefont{Blatter}},
  \bibinfo{journal}{Phys. Rev. Lett.} \textbf{\bibinfo{volume}{91}},
  \bibinfo{pages}{130404} (\bibinfo{year}{2003}).

\bibitem[{\citenamefont{Mathey et~al.}(2004)\citenamefont{Mathey, Wang,
  Hofstetter, Lukin, and Demler}}]{Mathey2004}
\bibinfo{author}{\bibfnamefont{L.}~\bibnamefont{Mathey}},
  \bibinfo{author}{\bibfnamefont{D.-W.} \bibnamefont{Wang}},
  \bibinfo{author}{\bibfnamefont{W.}~\bibnamefont{Hofstetter}},
  \bibinfo{author}{\bibfnamefont{M.~D.} \bibnamefont{Lukin}}, \bibnamefont{and}
  \bibinfo{author}{\bibfnamefont{E.}~\bibnamefont{Demler}},
  \bibinfo{journal}{Phys. Rev. Lett.} \textbf{\bibinfo{volume}{93}},
  \bibinfo{pages}{120404} (\bibinfo{year}{2004}).

\bibitem[{\citenamefont{Pollet et~al.}(2006)\citenamefont{Pollet, Troyer,
  Houcke, and Rombouts}}]{Pollet2006}
\bibinfo{author}{\bibfnamefont{L.}~\bibnamefont{Pollet}},
  \bibinfo{author}{\bibfnamefont{M.}~\bibnamefont{Troyer}},
  \bibinfo{author}{\bibfnamefont{K.} \bibnamefont{VanHoucke}},
  \bibnamefont{and} \bibinfo{author}{\bibfnamefont{S.~M.~A.}
  \bibnamefont{Rombouts}}, \bibinfo{journal}{Phys. Rev. Lett.}
  \textbf{\bibinfo{volume}{96}}, \bibinfo{eid}{190402}
  (\bibinfo{year}{2006}).

\bibitem[{\citenamefont{Pollet et~al.}(2008)\citenamefont{Pollet, Kollath,
  Schollwock, and Troyer}}]{Pollet2008}
\bibinfo{author}{\bibfnamefont{L.}~\bibnamefont{Pollet}},
  \bibinfo{author}{\bibfnamefont{C.}~\bibnamefont{Kollath}},
  \bibinfo{author}{\bibfnamefont{U.}~\bibnamefont{Schollwock}},
  \bibnamefont{and} \bibinfo{author}{\bibfnamefont{M.}~\bibnamefont{Troyer}},
  \bibinfo{journal}{Phys. Rev. A} \textbf{\bibinfo{volume}{77}},
  \bibinfo{eid}{023608} (\bibinfo{year}{2008}).

\bibitem[{\citenamefont{Hebert et~al.}(2008)\citenamefont{Hebert, Batrouni,
  Roy, and Rousseau}}]{Hebert2008}
\bibinfo{author}{\bibfnamefont{F.}~\bibnamefont{Hebert}},
  \bibinfo{author}{\bibfnamefont{G.~G.} \bibnamefont{Batrouni}},
  \bibinfo{author}{\bibfnamefont{X.}~\bibnamefont{Roy}}, \bibnamefont{and}
  \bibinfo{author}{\bibfnamefont{V.~G.} \bibnamefont{Rousseau}},
  \bibinfo{journal}{Phys. Rev. B} \textbf{\bibinfo{volume}{78}},
  \bibinfo{eid}{184505} (\bibinfo{year}{2008}).

\bibitem[{\citenamefont{Wen-Qiang~Ning and Lin}(2008)}]{null2008}
\bibinfo{author}{\bibnamefont{W.-Q~Ning}},
  \bibinfo{author}{\bibfnamefont{S-J~Gu}},
  \bibinfo{author}{\bibfnamefont{Y-G} \bibnamefont{Chen}},
  \bibinfo{author}{\bibfnamefont{C-Q} \bibnamefont{Wu}},
  \bibnamefont{and}
  \bibinfo{author}{\bibfnamefont{H-Q} \bibnamefont{Lin}},
  \bibinfo{journal}{J. Phys.: Condens. Matter} \textbf{\bibinfo{volume}{20}},
  \bibinfo{pages}{235236} (\bibinfo{year}{2008}).

\bibitem[{\citenamefont{Titvinidze et~al.}(2008)\citenamefont{Titvinidze,
  Snoek, and Hofstetter}}]{Titvinidze2008}
\bibinfo{author}{\bibfnamefont{I.}~\bibnamefont{Titvinidze}},
  \bibinfo{author}{\bibfnamefont{M.}~\bibnamefont{Snoek}}, \bibnamefont{and}
  \bibinfo{author}{\bibfnamefont{W.}~\bibnamefont{Hofstetter}},
  \bibinfo{journal}{Phys. Rev. Lett.} \textbf{\bibinfo{volume}{100}},
  \bibinfo{pages}{100401} (\bibinfo{year}{2008}).

\bibitem[{\citenamefont{Titvinidze et~al.}(2009)\citenamefont{Titvinidze,
  Snoek, and Hofstetter}}]{Titvinidze2009}
\bibinfo{author}{\bibfnamefont{I.}~\bibnamefont{Titvinidze}},
  \bibinfo{author}{\bibfnamefont{M.}~\bibnamefont{Snoek}}, \bibnamefont{and}
  \bibinfo{author}{\bibfnamefont{W.}~\bibnamefont{Hofstetter}},
  \bibinfo{journal}{Phys. Rev. B} \textbf{\bibinfo{volume}{79}},
  \bibinfo{pages}{144506} (\bibinfo{year}{2009}).

\bibitem[{\citenamefont{Orth et~al.}(2009)\citenamefont{Orth, Bergman, and
  Hur}}]{Orth2009}
\bibinfo{author}{\bibfnamefont{P.~P.} \bibnamefont{Orth}},
  \bibinfo{author}{\bibfnamefont{D.~L.} \bibnamefont{Bergman}},
  \bibnamefont{and} \bibinfo{author}{\bibfnamefont{K.} \bibnamefont{LeHur}},
  \bibinfo{journal}{Phys. Rev. A} \textbf{\bibinfo{volume}{80}},
  \bibinfo{eid}{023624} (\bibinfo{year}{2009}).

\bibitem[{\citenamefont{Orignac et~al.}(2010)\citenamefont{Orignac, Tsuchiizu,
  and Suzumura}}]{Orignac2010}
\bibinfo{author}{\bibfnamefont{E.}~\bibnamefont{Orignac}},
  \bibinfo{author}{\bibfnamefont{M.}~\bibnamefont{Tsuchiizu}},
  \bibnamefont{and} \bibinfo{author}{\bibfnamefont{Y.}~\bibnamefont{Suzumura}},
  \bibinfo{journal}{Phys. Rev. A} \textbf{\bibinfo{volume}{81}},
  \bibinfo{pages}{053626} (\bibinfo{year}{2010}).

\bibitem[{\citenamefont{Stewart et~al.}(2008)\citenamefont{Stewart, Gaebler,
  and Jin}}]{Stewart2008}
\bibinfo{author}{\bibfnamefont{J.~T.} \bibnamefont{Stewart}},
  \bibinfo{author}{\bibfnamefont{J.~P.} \bibnamefont{Gaebler}},
  \bibnamefont{and} \bibinfo{author}{\bibfnamefont{D.~S.} \bibnamefont{Jin}},
  \bibinfo{journal}{Nature} \textbf{\bibinfo{volume}{454}},
  \bibinfo{pages}{744} (\bibinfo{year}{2008}).

\bibitem[{\citenamefont{Gaebler et~al.}(2010)\citenamefont{Gaebler, Stewart,
  Drake, Jin, Perali, Pieri, and Strinati}}]{Gaebler2010}
\bibinfo{author}{\bibfnamefont{J.~P.} \bibnamefont{Gaebler}},
  \bibinfo{author}{\bibfnamefont{J.~T.} \bibnamefont{Stewart}},
  \bibinfo{author}{\bibfnamefont{T.~E.} \bibnamefont{Drake}},
  \bibinfo{author}{\bibfnamefont{D.~S.} \bibnamefont{Jin}},
  \bibinfo{author}{\bibfnamefont{A.}~\bibnamefont{Perali}},
  \bibinfo{author}{\bibfnamefont{P.}~\bibnamefont{Pieri}}, \bibnamefont{and}
  \bibinfo{author}{\bibfnamefont{G.~C.} \bibnamefont{Strinati}},
  \bibinfo{journal}{Nature Phys.} \textbf{\bibinfo{volume}{6}},
  \bibinfo{pages}{569} (\bibinfo{year}{2010}).

\bibitem[{\citenamefont{Perali et~al.}(2011)\citenamefont{Perali, Palestini,
  Pieri, Strinati, Stewart, Gaebler, Drake, and Jin}}]{Perali2011}
\bibinfo{author}{\bibfnamefont{A.}~\bibnamefont{Perali}},
  \bibinfo{author}{\bibfnamefont{F.}~\bibnamefont{Palestini}},
  \bibinfo{author}{\bibfnamefont{P.}~\bibnamefont{Pieri}},
  \bibinfo{author}{\bibfnamefont{G.~C.} \bibnamefont{Strinati}},
  \bibinfo{author}{\bibfnamefont{J.~T.} \bibnamefont{Stewart}},
  \bibinfo{author}{\bibfnamefont{J.~P.} \bibnamefont{Gaebler}},
  \bibinfo{author}{\bibfnamefont{T.~E.} \bibnamefont{Drake}}, \bibnamefont{and}
  \bibinfo{author}{\bibfnamefont{D.~S.} \bibnamefont{Jin}},
  \bibinfo{journal}{Phys. Rev. Lett.} \textbf{\bibinfo{volume}{106}},
  \bibinfo{pages}{060402} (\bibinfo{year}{2011}).

\bibitem[{\citenamefont{Schirotzek et~al.}(2009)\citenamefont{Schirotzek, Wu,
  Sommer, and Zwierlein}}]{Schirotzek2009}
\bibinfo{author}{\bibfnamefont{A.}~\bibnamefont{Schirotzek}},
  \bibinfo{author}{\bibfnamefont{C.-H.} \bibnamefont{Wu}},
  \bibinfo{author}{\bibfnamefont{A.}~\bibnamefont{Sommer}}, \bibnamefont{and}
  \bibinfo{author}{\bibfnamefont{M.~W.} \bibnamefont{Zwierlein}},
  \bibinfo{journal}{Phys. Rev. Lett.} \textbf{\bibinfo{volume}{102}},
  \bibinfo{pages}{230402} (\bibinfo{year}{2009}).

\bibitem[{\citenamefont{Nascimb\`ene et~al.}(2009)\citenamefont{Nascimb\`ene,
  Navon, Jiang, Tarruell, Teichmann, McKeever, Chevy, and
  Salomon}}]{Nascimb`ene2009}
\bibinfo{author}{\bibfnamefont{S.}~\bibnamefont{Nascimb\`ene}},
  \bibinfo{author}{\bibfnamefont{N.}~\bibnamefont{Navon}},
  \bibinfo{author}{\bibfnamefont{K.~J.} \bibnamefont{Jiang}},
  \bibinfo{author}{\bibfnamefont{L.}~\bibnamefont{Tarruell}},
  \bibinfo{author}{\bibfnamefont{M.}~\bibnamefont{Teichmann}},
  \bibinfo{author}{\bibfnamefont{J.}~\bibnamefont{McKeever}},
  \bibinfo{author}{\bibfnamefont{F.}~\bibnamefont{Chevy}}, \bibnamefont{and}
  \bibinfo{author}{\bibfnamefont{C.}~\bibnamefont{Salomon}},
  \bibinfo{journal}{Phys. Rev. Lett.} \textbf{\bibinfo{volume}{103}},
  \bibinfo{pages}{170402} (\bibinfo{year}{2009}).

\bibitem[{\citenamefont{Byczuk and Vollhardt}(2009)}]{Byczuk2009b}
\bibinfo{author}{\bibfnamefont{K.}~\bibnamefont{Byczuk}} \bibnamefont{and}
  \bibinfo{author}{\bibfnamefont{D.}~\bibnamefont{Vollhardt}},
  \bibinfo{journal}{Ann. Phys.} \textbf{\bibinfo{volume}{18}},
  \bibinfo{pages}{622} (\bibinfo{year}{2009}).

\bibitem[{\citenamefont{Metzner and Vollhardt}(1989)}]{Metzner1989}
\bibinfo{author}{\bibfnamefont{W.}~\bibnamefont{Metzner}} \bibnamefont{and}
  \bibinfo{author}{\bibfnamefont{D.}~\bibnamefont{Vollhardt}},
  \bibinfo{journal}{Phys. Rev. Lett.} \textbf{\bibinfo{volume}{62}},
  \bibinfo{pages}{324} (\bibinfo{year}{1989}).

\bibitem[{\citenamefont{Georges et~al.}(1996)\citenamefont{Georges, Kotliar,
  Krauth, and Rozenberg}}]{Georges1996}
\bibinfo{author}{\bibfnamefont{A.}~\bibnamefont{Georges}},
  \bibinfo{author}{\bibfnamefont{G.}~\bibnamefont{Kotliar}},
  \bibinfo{author}{\bibfnamefont{W.}~\bibnamefont{Krauth}}, \bibnamefont{and}
  \bibinfo{author}{\bibfnamefont{M.~J.} \bibnamefont{Rozenberg}},
  \bibinfo{journal}{Rev. Mod. Phys.} \textbf{\bibinfo{volume}{68}},
  \bibinfo{pages}{13} (\bibinfo{year}{1996}).

\bibitem[{\citenamefont{Wilson}(1975)}]{Wilson1975}
\bibinfo{author}{\bibfnamefont{K.~G.} \bibnamefont{Wilson}},
  \bibinfo{journal}{Rev. Mod. Phys.} \textbf{\bibinfo{volume}{47}},
  \bibinfo{pages}{773} (\bibinfo{year}{1975}).

\bibitem[{\citenamefont{Bulla et~al.}(2008)\citenamefont{Bulla, Costi, and
  Pruschke}}]{Bulla2008}
\bibinfo{author}{\bibfnamefont{R.}~\bibnamefont{Bulla}},
  \bibinfo{author}{\bibfnamefont{T.~A.} \bibnamefont{Costi}}, \bibnamefont{and}
  \bibinfo{author}{\bibfnamefont{T.}~\bibnamefont{Pruschke}},
  \bibinfo{journal}{Rev. Mod. Phys.} \textbf{\bibinfo{volume}{80}},
  \bibinfo{pages}{395} (\bibinfo{year}{2008}).

\end{thebibliography}

\end{document}